\documentclass[12pt]{iopart}

\usepackage{graphicx}
\begin{document}

\title[A simple route to a tunable electromagnetic gateway]{A simple route to a tunable electromagnetic gateway}

\author{Huanyang Chen and Che Ting Chan}

\address{Department of Physics and the William Mong Institute of NanoScience and
Technology, The Hong Kong University of Science and Technology,
Clear Water Bay, Hong Kong, China}

\author{Shiyang Liu and Zhifang Lin}

\address{Surface Physics Laboratory, Department of Physics, Fudan University,
Shanghai 200433, China}

\begin{abstract}
Transformation optics is used to design a gateway that can block
electromagnetic waves but allows the passage of other entities. Our
conceptual device has the advantage that it can be realized with
simple materials and structural parameters and can have a reasonably
wide bandwidth. In particular, we show that our system can be
implemented by using a magnetic photonic crystal structure that
employs a square ray of ferrite rods, and as the field response of
ferrites can be tuned by external magnetic fields, we end up with an
electromagnetic gateway that can be open or shut using external
fields. The functionality is also robust against the positional
disorder of the rods that made up the photonic crystal.
\end{abstract}

\pacs{41.20.Jb, 42.25.Fx, 42.25.Gy}

\maketitle

\section{Introduction}

Transformation optics \cite{Leonhardt:2006, Pendry:2006,
Greenleaf:2003} has paved the way for the development of optical
devices that can realize functionalities that were thought to be
possible only in science fictions
\cite{Schurig:2006}-\cite{Lai:2009}. One such conceptual device that
has attracted great public interest is a gateway that can block
electromagnetic waves but that allows the passage of other entities.
This device can be viewed as an implementation of a ``hidden
portal'' mentioned in fictions \cite{Rowling:1998, Luo:1}. However,
the feasibility of such devices is limited by the very complex
material parameters and the narrow bandwidth. Here, we show that
gateway-type devices can actually be realized with simple parameters
and they can have wider band widths such that the concept is closer
to reality \cite{Shelby:2001}-\cite{Valentine:2008} than previously
thought. The structure can be implemented by using the magnetic
photonic crystal structures that are field tunable, resulting in an
invisible electromagnetic gateway that can be open or shut using
magnetic fields.

\section{A new ``superscatterer''}

\begin{figure}
\begin{center}
\includegraphics[angle=-0,width=0.75\columnwidth] {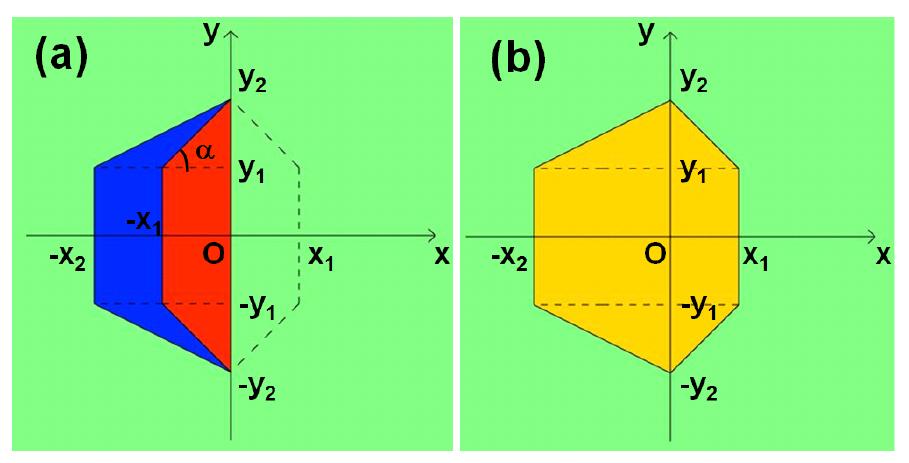}
\end{center}
\caption{\textbf{A simple electromagnetic structure comprising of
two homogenous regions (left panel) and an equivalent structure that
gives the same scattering (right panel).} (a) The schematic plot for
the scatterer that appears electromagnetically larger than its
physical size. The red region is the DNM, the blue region is the
original isotropic material that is to the left of the DNM, and the
green region represents the air background. $x_1 $, $x_2 $, $y_1 $,
$y_2 $ and $\alpha $ are the detailed coordinates of the structure.
The dashed lines in the region ($x>0$) give the virtual boundary for
the equivalent transformation medium. (b) The equivalent
transformation medium is indicated by the yellow region, whose
parameters are implemented by equation (\ref{eq2}). The dashed lines
distinguish the three different regions for the definition of
$p$.}\label{fig.1}
\end{figure}

We start from a simple implementation of transformation optics. In
Fig. \ref{fig.1}(a), an object (colored in blue) is placed to the
left side of a double negative medium (DNM) with $\varepsilon =\mu =
-1$. Let the object be of permittivity $\varepsilon _0 $ and
permeability $\mu _0 $. The detailed shape and the length scale of
the structure are shown in Fig. \ref{fig.1}(a). From the viewpoint
of transformation optics, the whole structure is optically
equivalent to another object described in Fig. \ref{fig.1}(b) for
far field observers. The equivalent permittivity and permeability
tensors are
$\mathord{\buildrel{\lower3pt\hbox{$\scriptscriptstyle\leftrightarrow$}}\over
{\varepsilon }} = \varepsilon _0
\mathord{\buildrel{\lower3pt\hbox{$\scriptscriptstyle\leftrightarrow$}}\over
{c}} $ and
$\mathord{\buildrel{\lower3pt\hbox{$\scriptscriptstyle\leftrightarrow$}}\over
{\mu }} = \mu _0
\mathord{\buildrel{\lower3pt\hbox{$\scriptscriptstyle\leftrightarrow$}}\over
{c}} $ with a constant tensor
$\mathord{\buildrel{\lower3pt\hbox{$\scriptscriptstyle\leftrightarrow$}}\over
{c}} $. We introduce some parameters for ease of reference following
the coordinates $x_1 $, $x_2 $, $y_1 $, $y_2 $ and the angle $\alpha
$ in Fig. \ref{fig.1}. Let $\Delta = x_2 - x_1 $ be the waist of the
object in the $x$-direction, $r = \frac{\Delta }{2x_1 + \Delta }$ be
a coordinate compression ratio and

\begin{equation}
\label{eq1} p = \left\{ {{\begin{array}{*{20}c}
 {\frac{1}{\tan \alpha }\frac{2(x_1 + \Delta )}{2x_1 + \Delta },\mbox{ }y_1
< y < y_2 ,\mbox{ } - x_2 \frac{y_2 - y}{y_2 - y_1 } < x < x_1
\frac{y_2 - y}{y_2 -
y_1 },} \hfill \\
 {0,\mbox{ }\vert y\vert < y_1 ,\mbox{ } - x_2 < x < x_1 ,} \hfill \\
 { - \frac{1}{\tan \alpha }\frac{2(x_1 + \Delta )}{2x_1 + \Delta },\mbox{ }
- y_2 < y < - y_1 ,\mbox{ } - x_2 \frac{y_2 + y}{y_2 - y_1 } < x <
x_1 \frac{y_2 +
y}{y_2 - y_1 },} \hfill \\
\end{array} }} \right.
\end{equation}

\noindent where $(x,\;y)$ are the position coordinates of the
equivalent transformation medium in Fig. \ref{fig.1}(b). The
inequalities for the definition of the parameter, $p$, come from
three different coordinate transformations and the three regions are
distinguished by the two dashed lines in Fig. \ref{fig.1}(b).

It can be proved that (see Appendix A):

\begin{equation}
\label{eq2}
\mathord{\buildrel{\lower3pt\hbox{$\scriptscriptstyle\leftrightarrow$}}\over
{c}} = \left[ {{\begin{array}{*{20}c}
 {\frac{1 + p^2}{r}} \hfill & { - p} \hfill & 0 \hfill \\
 { - p} \hfill & r \hfill & 0 \hfill \\
 0 \hfill & 0 \hfill & r \hfill \\
\end{array} }} \right].
\end{equation}

The coordinate transformation is based on the recent work on
transformation optics with complementary medium (or folded
geometries) \cite{Leonhardt:2007, Pendry:2003}. The original object
is mapped into the equivalent transformation medium. The DNM
together with its mirror image (the boundary is marked by the dashed
line in Fig. \ref{fig.1}(a)) of air form a pair of complementary
media, in the sense the phase accumulated in one segment is exactly
cancelled by another. A more detailed discussion of the optical
property of this specific geometry and the associated coordinate
transformation can be found in Appendix A. The structure in Fig.
\ref{fig.1}(a) can be treated as a form of a scattering amplifier
\cite{Yang:2008} because its scattering cross section can be much
larger than its geometric cross section. The amplification effect
originates from the excitation of surface plasmons in the DNM and
air interface.

\begin{figure}
\begin{center}
\includegraphics[angle=-0,width=0.75\columnwidth] {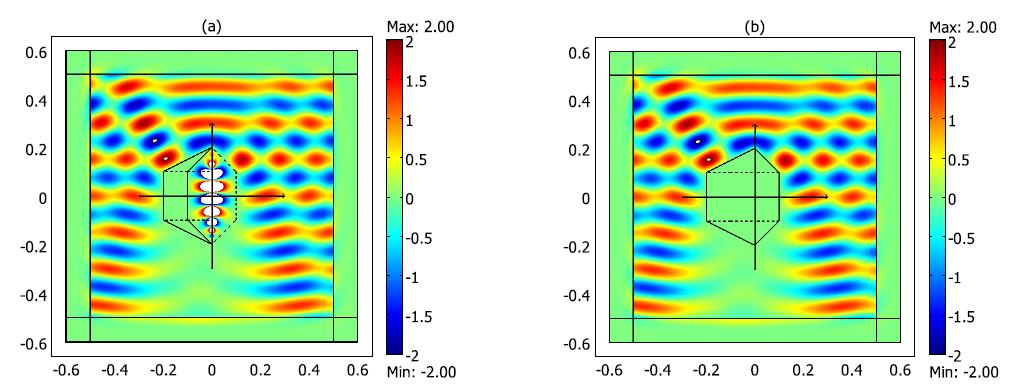}
\end{center}
\caption{\textbf{The scattering pattern for the scattering object
and the equivalent scatterer.} (a) The scattering pattern of the
structure depicted in Fig. \ref{fig.1}(a). (b) The scattering
pattern of the equivalent material but with the PEC replacing the
material implemented by equation (\ref{eq2}).}\label{fig.2}
\end{figure}

As a concrete example, we suppose that the original isotropic
material (the blue region in Fig. \ref{fig.1}(a)) has a large value
of permittivity, $\varepsilon _0 = - 10000$ and $\mu _0 = 1$, which
may be treated approximately as a perfect electric conductor (PEC).
Figure \ref{fig.2}(a) shows the scattering field pattern of the
structure in Fig. \ref{fig.1}(a). The plane wave is incident from
the top to the bottom and has transverse electric (TE) polarization,
for which the E field is along $z$-direction. In this paper, we only
consider the TE modes for simplicity. Note that the same idea works
for transverse magnetic (TM) modes as well. The frequency is $2 GHz$
and $x_1 = y_1 = 0.1\;m$, $x_2 = y_2 = 0.2\;m$ and $\alpha = \pi /
4$. The structure behaves like an equivalent material with
parameters described by equation (\ref{eq2}) to the far-field
observers. Due to the large mismatch of the impedance of the
equivalent material with the air background, we expect that the
equivalent object (the yellow domain in Fig. \ref{fig.1}(b)) should
scatter like a PEC, and for that reason we choose for comparison in
Fig. \ref{fig.2}(b) the scattering field pattern of a perfect
conductor filling up the entire domain of the virtual object in Fig.
\ref{fig.1}(b). The similar far-field scattered patterns between
Fig. \ref{fig.2}(a) and Fig. \ref{fig.2}(b) confirm the strong
scattering effect. The permittivity and permeability of the DNM are
actually chosen as $-1+0.0001\times i$ in the simulations to avoid
numerical divergence problems \cite{Pendry:2003}. The PEC-like
scatters are always chosen as the materials of permittivity at
$-10000$ and the permeability is taken to be $1$ in this paper for
simplicity.

\section{An invisible electromagnetic gateway}

\begin{figure}
\begin{center}
\includegraphics[angle=-0,width=0.60\columnwidth] {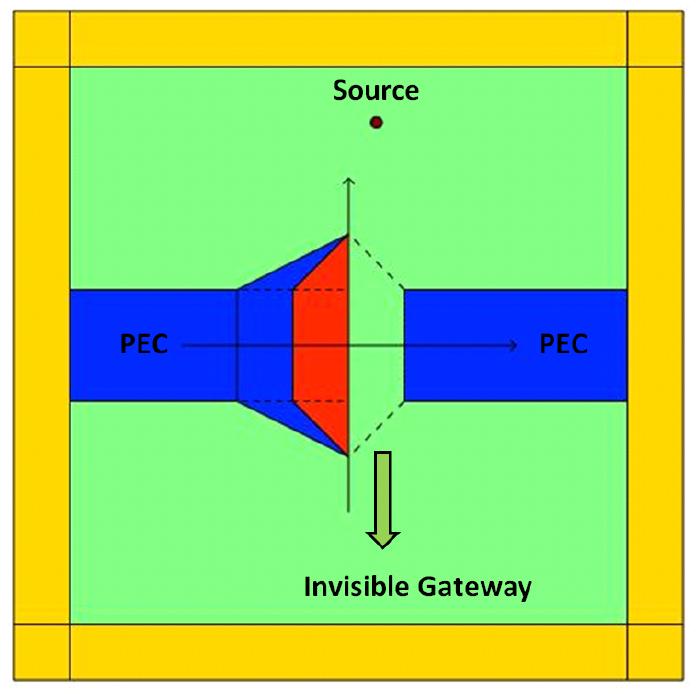}
\end{center}
\caption{\textbf{The computation domains for the electromagnetic
gateway.} The yellow regions are the perfectly matched layers
(PMLs), the red region is the DNM, the green region is air, the blue
regions are PECs. The point source is located at above the gateway
and it is denoted by a brown circle. The region between the DNM and
the right PEC is the so called ``invisible gateway''.}\label{fig.3}
\end{figure}

\begin{figure}
\begin{center}
\includegraphics[angle=-0,width=0.90\columnwidth] {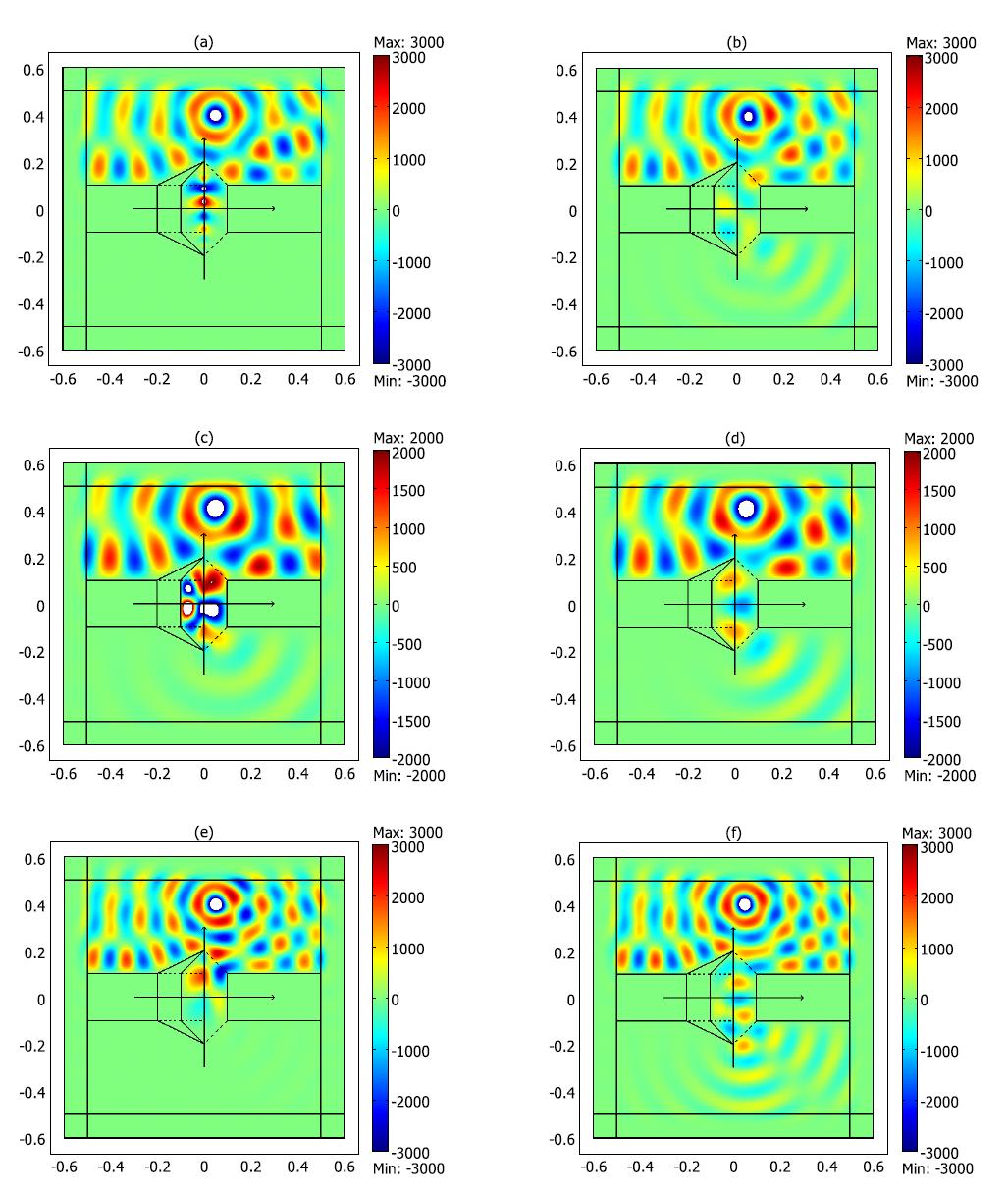}
\end{center}
\caption{\textbf{The functionalities of the gateway (with or without
the DNM) at different frequencies.} (a) The electric field pattern
for the gateway at $2 GHz$. (b) The electric field pattern for the
gateway without the DNM at $2 GHz$. (c) Same as (a) but for $1.6
GHz$. (d) Same as (b) but for $1.6 GHz$. (e) Same as (a) but for
$2.4 GHz$. (f) Same as (b) but for $2.4 GHz$. }\label{fig.4}
\end{figure}

The amplified scattering effect can be utilized to make an invisible
gateway \cite{Luo:1}. Suppose that a PEC wall separates the whole
space into two regions, the upper domain and the lower domain. If
there are channels (or gateways) opened in the PEC wall, people in
the two different spaces can communicate with each other, both
physically and through EM waves. However, if we replace the doors
with the above-described configuration at a specific frequency, the
communication for that frequency will be blocked because the systems
behave like PECs. The most amazing fact is that the channel is in
fact physically empty. There is nothing but air in the channel so
objects can ``walk through'' but the channel is blocked as perceived
by the eye because light at the designated frequency cannot
penetrate. Figure \ref{fig.3} is a schematic plot of such a gateway
based on the above idea. With the same scale as the one in Fig.
\ref{fig.2}(a), we demonstrate the properties of such a gateway. We
suppose that there is a line source located at $(0.05 m, 0.4 m)$
with a frequency of $2 GHz$ in the upper domain. Figure
\ref{fig.4}(a) shows that the waves cannot pass through the gateway
and are excluded from the lower domain. However, without the DNM,
the waves can propagate into the lower region as shown in Fig.
\ref{fig.4}(b).

We now consider the band-width of such a device. To be more
concrete, we consider the following dispersions for the DNM:

\begin{equation}
\label{eq3}
\begin{array}{l}
 \varepsilon = 1 - \frac{f_p^2 }{f(f + i\Gamma )}, \\
 \mu = 1 - \frac{Ff^2}{f^2 - f_0^2 + i\gamma f}, \\
 \end{array}
\end{equation}

\noindent where $f_p = 2.828\;GHz$, $\Gamma = 0.1\;MHz$, $F = 1.5$,
$f_0 = 1\;GHz$ and $\gamma = 0.075\;MHz$. When $f = 2\;GHz$, the
relative permittivity and permeability return to $-1+0.0001\times i$
used in the above simulations. When $f = 1.6\;GHz$, the impedances
do not match at the interface of the air and the DNM while the
refractive index of the DNM is about $-1.76$. Figure \ref{fig.4}(c)
shows the electric field pattern while Fig. \ref{fig.4}(d) shows the
field without the DNM. The DNM can reduce the penetration of the
waves from the upper region when compared with Fig. \ref{fig.4}(c)
and Fig. \ref{fig.4}(d). However, as $n<-1$, there is still a
``slit'' between the virtual boundary and the right PEC, which
allows the waves to propagate somewhat into the lower region. If
$n=-1$, the virtual boundary is simply the mirror of the interface
of the DNM and PEC on the left with $x=0$ as the mirror. The
position of the virtual boundary can be obtained heuristically from
the image-forming principle. When $f = 2.4\;GHz$, the impedances are
still mismatched while the refractive index of the DNM is about
$-0.56$. As the absolute value of $n$ decreases, the virtual
boundary of the image expands and there will be no passage for the
waves to penetrate because the virtual boundary overlaps with the
PEC on the right-band side. For example, we plot the electric field
pattern in Fig. \ref{fig.4}(e) and the case without the DNM in Fig.
\ref{fig.4}(f). We find that the DNM can eliminate the penetration
of the waves from the upper region. That means that the negative
band of the DNM beyond $f = 2\;GHz$ (or $-1<n<0$) is the working
frequency of the designed gateway. However, we also find that the
wave blocking effect will become weaker with higher frequencies (in
this case, about $2.5 GHz$) in our simulations. As negative index
media are intrinsically dispersive, the bandwidth has to be finite
as the parameters will deviate progressively from those required by
transformation optics. But the present gateway is shown to be
relatively robust and has a broad operation bandwidth of about
$20{\%}$, which can be regarded as a broadband device. To have a
broader bandwidth, we can simply reduce the distance of the DNM and
the right PEC to enhance the overlapping while the functionality of
the gateway is sacrificed. As there are extensive designs of the
DNMs at various wavelengths (both theoretically and experimentally)
\cite{Shelby:2001}-\cite{Valentine:2008}, it would be reasonably
feasible for the present gateway to be realized in the future.

If the losses of the metamaterial are large (i.e., the imaginary
parts of the parameters of the DNM are larger than $0.1$), the
bandwidth of the gateway will be small and the functionality will be
compromised. The absorption of the DNM is the key difficulty for
both the gateway described here as for other devices such as the
perfect lens. All our simulation results are calculated using the
COMSOL Multiphysics finite element-based electromagnetics solver.

\section{The implementation of a tunable electromagnetic gateway}

Very recently, it has been demonstrated that the DNM can be realized
with a simple array of ferrite rods without any metallic components
\cite{Liu:2008}, which can be used to implement the present gateway.
One of the unique merits of this magnetic metamaterial is that its
optical properties are magnetically tunable. As such, the gateway
can be manipulated using external magnetic fields. The magnetic
metamaterial is a periodic square array of subwavelength ferrite
rods with radii of $r = 3.5 mm$ and a lattice constant of $a = 10
mm$. The permittivity is taken to be $\varepsilon = 25$, and the
permeability of the ferrite rods has the form $\hat {\mu } = \left(
{{\begin{array}{*{20}c}
 {\mu _r } \hfill & { - i\mu _\kappa } \hfill & 0 \hfill \\
 {i\mu _\kappa } \hfill & {\mu _r } \hfill & 0 \hfill \\
 0 \hfill & 0 \hfill & 1 \hfill \\
\end{array} }} \right)$ with $\mu _r = 1 + \frac{\omega _m (\omega _0 -
2\pi i\alpha_{1} f )}{(\omega _0 - 2\pi i\alpha_{1} f )^2 - 4 \pi ^2
f^2}$ and $\mu _\kappa = \frac{2\pi \omega _m f }{(\omega _0 - 2\pi
i\alpha_{1} f )^2 - 4\pi ^2 f ^2}$, where $\alpha_{1} $ is the
damping coefficient, $\omega _0 = \gamma_{1} H_0 $ is the resonance
frequency with $\gamma_{1} $ being the gyromagnetic ratio; $H_0 $ is
the sum of the external magnetic field and shape anisotropy field
along the $z$-direction, $\omega _m = \gamma_{1} M_s $ is the
characteristic frequency with $M_s $ the saturation magnetization
and it is taken as $M_s = 1750 G$, typical for single-crystal
yttrium-iron-garnet (YIG). As the absorption of single-crystal YIG
is extremely low, we can set $\alpha_{1} = 0$.

\begin{figure}
\begin{center}
\includegraphics[angle=-0,width=0.75\columnwidth] {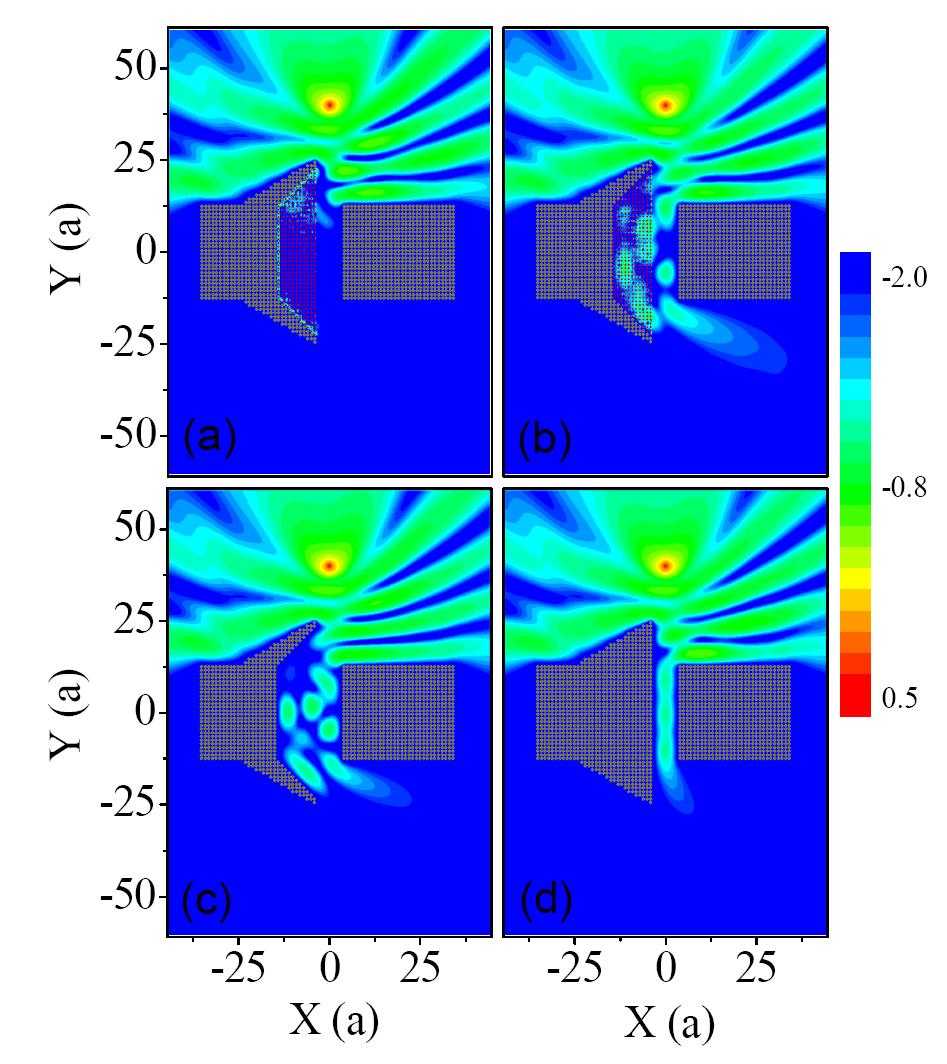}
\end{center}
\caption{\textbf{The implementation of the invisible gateway using
magnetic photonic crystals. Here, we show the electric field
intensity pattern for different geometries working at different
values of the external magnetic field.} (a) $H_0 = 500 Oe$, (b) $H_0
= 475 Oe$, (c) without magnetic metamaterials, (d) with magnetic
metamaterial replaced by PEC.}\label{fig.5}
\end{figure}

The numerical simulations on the gateway under different conditions
are performed by using the multiple scattering method
\cite{Liu:2008}. Figure \ref{fig.5} shows the electric field
intensity in the logarithmic scale. The bulk PEC is replaced by a
discrete system of the periodically arranged PEC rods (marked with
dark green solid circles in the figure), and the ferrite rods are
denoted as hollow circles. Under $H_0 = 500 Oe$ and the frequency $f
= 2.55\;GHz$, the effective refractive index of the magnetic
metamaterial is $n = - 1(\varepsilon _{eff} = \mu _{eff} = - 1)$. In
Fig. \ref{fig.5}(a), it can be seen that the electric field is
excluded from the air channel, so that the open channel appears to
be blocked to the eye at this frequency. This demonstrates that the
gateway can be implemented by a photonic crystal type structure in
which each element (each ferrite rod in the periodic array) is
identical. By changing the external field to $H_0 = 475 Oe$, the
refractive index of the magnetic metamaterial becomes $n = 1$(
$\varepsilon _{eff} = \mu _{eff} = 1)$. The electric field intensity
under such conditions is shown in Fig. \ref{fig.5}(b), which shows
that the channel is open for EM wave passage, or the passage appears
to be open to the eye. Our results demonstrate that the gateway can
be implemented by a very simple configuration and the effect is
tunable by an external field. As shown in Appendix B, the effect is
fairly robust to the disorder of the position of ferrite cylinders.
Figures \ref{fig.5}(c) and \ref{fig.5}(d) illustrate the fact that
the channel is electromagnetically open if the ferrite rods are
replaced by air and by PEC rods. In addition to the simple geometry
and weak absorption, the magnetic metamaterial has a reasonably
broad bandwidth as a DNM.

\section{Conclusion}

In conclusion, we showed that a robust and tunable electromagnetic
gateway can be constructed using simple material parameters. The
idea is conceived through consideration of transformation optics and
can be realized using a photonic crystal type structure.

\ack
%\section*{Acknowledgments}
We thank Dr. Yun Lai and Dr. Jack Ng for their helpful discussions.
This work was supported by Hong Kong Central Allocation Grant No.
HKUST3/06C. Computation resources were supported by the Shun Hing
Education and Charity Fund. S.Y.L and Z.F.L were supported by
CNKBRSF, NNSFC, PCSIRT, MOE of China (B06011), and Shanghai Science
and Technology Commission.

\section*{Appendix A}
\appendix
\setcounter{section}{1}

\noindent In this appendix, we give the detailed coordinate
transformation to produce the folded geometry used in the text. Let
us first consider the following coordinate transformation,

\begin{equation}
\label{eqa1}
\begin{array}{l}
 x = \left\{ {{\begin{array}{*{20}c}
 { - x_2^y + \frac{x_2 + x_1 }{x_2 - x_1 }\times (x' + x_2^y ),\quad - x_2^y
< x' < - x_1^y ,} \hfill \\
 { - x',\mbox{ } - x_1^y < x' < 0,} \hfill \\
 {x',\mbox{ }else,} \hfill \\
\end{array} }} \right. \\
 y' = y,\quad z' = z, \\
 \end{array}
\end{equation}

\noindent where

\begin{equation}
\label{eqa2} (x_1^y ,\;x_2^y ) = \left\{ {{\begin{array}{*{20}c}
 {(x_1 \frac{y_2 - y}{y_2 - y_1 },\;x_2 \frac{y_2 - y}{y_2 - y_1 }),\mbox{
}y_1 < y < y_2 } \hfill \\
 {(x_1 ,\;x_2 ),\mbox{ } - y_1 < y < y_1 } \hfill \\
 {(x_1 \frac{y_2 + y}{y_2 - y_1 },\;x_2 \frac{y_2 + y}{y_2 - y_1 }),\mbox{ }
- y_2 < y < - y_1 } \hfill \\
\end{array} }} \right.,
\end{equation}

This coordinate transformation maps the virtual space ($x$-space)
described in Fig. \ref{fig.1}(b) into the physical space
($x'$-space) described in Fig. \ref{fig.1}(a). If the virtual space
is a vacuum, we can obtain the corresponding material parameters in
the physical space. It contains two kinds of materials in the vacuum
background, one a double negative medium (DNM) with $\varepsilon =
\mu = - 1$ (shown in red in Fig. \ref{fig.1}(a)), the other an
anisotropic medium (shown in blue in Fig. \ref{fig.1}(a)) whose
parameters are derived in the following section.

We can easily obtain the Jacobian transformation matrix,

\begin{equation}
\label{eqa3} \Lambda = \left[ {{\begin{array}{*{20}c}
 {\frac{\partial x'}{\partial x}} \hfill & {\frac{\partial x'}{\partial y}}
\hfill & 0 \hfill \\
 0 \hfill & 1 \hfill & 0 \hfill \\
 0 \hfill & 0 \hfill & 1 \hfill \\
\end{array} }} \right]
\end{equation}

\noindent from equations (\ref{eqa1}) and (\ref{eqa2}) for the
mapping between the blue region in Fig. \ref{fig.1}(a) and the
yellow region in Fig. \ref{fig.1}(b), and we can also prove that
$\frac{\partial x'}{\partial x} = r$ and $\frac{\partial
x'}{\partial y} = p$ in the text. From transformation optics, we can
obtain the material parameters for the blue region,

\begin{equation}
\label{eqa4}
\mathord{\buildrel{\lower3pt\hbox{$\scriptscriptstyle\leftrightarrow$}}\over
{\varepsilon }} _p =
\mathord{\buildrel{\lower3pt\hbox{$\scriptscriptstyle\leftrightarrow$}}\over
{\mu }} _p = \left[ {{\begin{array}{*{20}c}
 {\frac{r^2 + p^2}{r}} \hfill & {\frac{p}{r}} \hfill & 0 \hfill \\
 {\frac{p}{r}} \hfill & {\frac{1}{r}} \hfill & 0 \hfill \\
 0 \hfill & 0 \hfill & {\frac{1}{r}} \hfill \\
\end{array} }} \right].
\end{equation}

\begin{figure}
\begin{center}
\includegraphics[angle=-0,width=0.90\columnwidth] {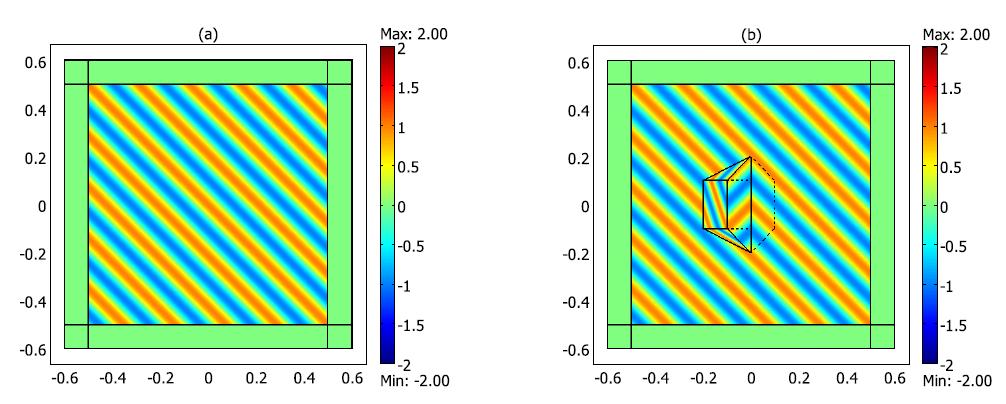}
\end{center}
\caption{\textbf{The folded geometry used in the text.} (a)A
obliquely incident TE plane wave in free space. (b)The structure
with the material described by equation (\ref{eqa4}) and the DNM are
invisible.}\label{fig.a1}
\end{figure}

Figure \ref{fig.a1}(a) shows the propagation of the incident
transverse electric (TE) polarization plane wave from the left to
the right but at an angle of 45 degrees from the $x$-direction in
free space. The frequency is $2 GHz$. We consider the object
described in equation (\ref{eqa4}) together with an DNM domian. The
scales of the structure in our simulations are the same as the ones
used in the text. Figure \ref{fig.a1}(b) shows the scattering
pattern of such a structure and demonstrates that the structure
itself is invisible to the far-field observers.

\begin{figure}
\begin{center}
\includegraphics[angle=-0,width=0.90\columnwidth] {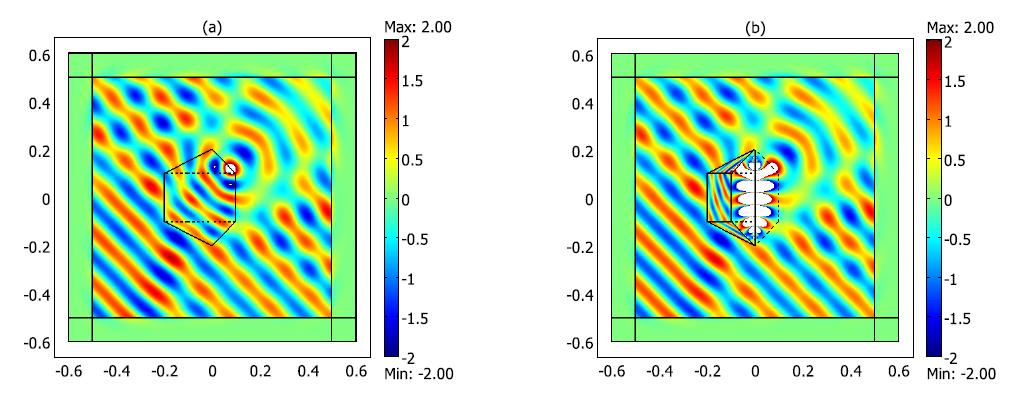}
\end{center}
\caption{\textbf{The scattering pattern of an equivalent scatterer
and the real scatterer.} (a) The scattering pattern of an object of
permittivity $\varepsilon _s = 2.25$. (b) The corresponding
scattering pattern of the scattering amplifier.}\label{fig.a2}
\end{figure}

A modification of the above folded geometry can lead to a variety of
conceptual devices. For example, if the virtual space is not a
vacuum but an object of permittivity $\varepsilon _s $ and
permeability $\mu _s $, then the parameters in the blue region
should be a permittivity tensor, $\varepsilon _s
\mathord{\buildrel{\lower3pt\hbox{$\scriptscriptstyle\leftrightarrow$}}\over
{\varepsilon }} _p $, and a permeability tensor, $\mu _s
\mathord{\buildrel{\lower3pt\hbox{$\scriptscriptstyle\leftrightarrow$}}\over
{\mu }} _p $. The object in the blue region together with the DNM
domain will have a scattering cross section that can be much larger
than its geometric cross section. Figure \ref{fig.a2}(a) shows the
scattering pattern of an object with permittivity $\varepsilon _s =
2.25$ and permeability $\mu _s = 1$. Figure \ref{fig.a2}(b) shows
the scattering pattern of the structure including the object with
permittivity tensor $\varepsilon _s
\mathord{\buildrel{\lower3pt\hbox{$\scriptscriptstyle\leftrightarrow$}}\over
{\varepsilon }} _p $ and permeability tensor $\mu _s
\mathord{\buildrel{\lower3pt\hbox{$\scriptscriptstyle\leftrightarrow$}}\over
{\mu }} _p $ and the DNM. The two similar far-field patterns in
Figs. \ref{fig.a2}(a) and \ref{fig.a2}(b) confirm the enhanced
scattering effect. In the text, we apply this structure in an
inverse manner. We suppose that the object in the blue region is of
permittivity $\varepsilon _0 $ and permeability $\mu _0 $. Then, the
whole structure is equivalent to another object with its
permittivity and permeability tensors,
$\mathord{\buildrel{\lower3pt\hbox{$\scriptscriptstyle\leftrightarrow$}}\over
{\varepsilon }} = \varepsilon _0
\mathord{\buildrel{\lower3pt\hbox{$\scriptscriptstyle\leftrightarrow$}}\over
{c}} $ and
$\mathord{\buildrel{\lower3pt\hbox{$\scriptscriptstyle\leftrightarrow$}}\over
{\mu }} = \mu _0
\mathord{\buildrel{\lower3pt\hbox{$\scriptscriptstyle\leftrightarrow$}}\over
{c}} $ (described in the text). We can easily obtain equation
(\ref{eq2}) by performing an inverse transformation.

\section*{Appendix B}
\appendix
\setcounter{section}{2}

\begin{figure}
\begin{center}
\includegraphics[angle=-0,width=0.75\columnwidth] {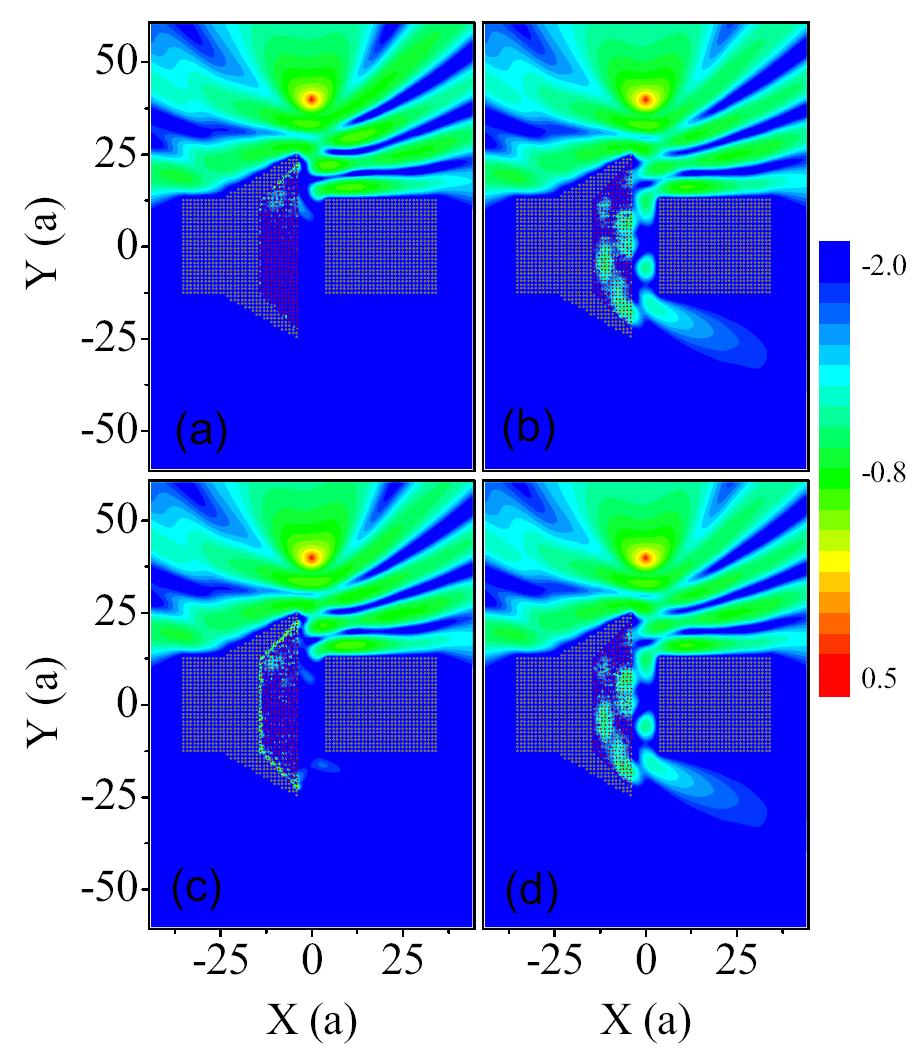}
\end{center}
\caption{\textbf{The robust effect from the disorder. The electric
field intensity pattern for the gateway with different levels of
disorder working at different external magnetic fields.} (a) $s =
0.5$, $H_0 = 500\;Oe$; (b) $s = 0.5$, $H_0 = 475\;Oe$; (c) $s =
1.0$, $H_0 = 500\;Oe$; (d) $s = 1.0$, $H_0 =
475\;Oe$.}\label{fig.b1}
\end{figure}

In this appendix, we demonstrate the robustness for the
implementation of the gateway using the magnetic photonic crystal
structure. We introduce disorder by displacing the ferrite rods from
their original positions with

\begin{equation}
\label{eqb1} \Delta r = (a - 2r)\times s\times r_n ,
\end{equation}

\noindent where $a$ is the lattice constant, $r$ is the radii of the
ferrite rods, $r_n $ is a uniform random variable on the interval
$(-0.5, 0.5)$ (assures no overlap of cylinders), $s$ is the strength
of disorder. Figure \ref{fig.b1}(a) corresponds to Fig.
\ref{fig.5}(a) but with a disorder of $s = 0.5$ introduced. The
gateway remains blocked to the eye for this strength of disorder.
When $H_0 $ changes to $475\;Oe$, the gateway is open again as shown
in Fig. \ref{fig.b1}(b). Figure \ref{fig.b1}(c) corresponds to Fig.
\ref{fig.5}(a) but with a strong disorder of $s = 1.0$ introduced.
The functionality of the gateway is now slightly compromised but the
blocking is at least partially effective at such a strength of
disorder. With $H_0 = 475\;Oe$, the gateway is open again as shown
in Fig. \ref{fig.b1}(d). The results indicate that the functionality
is robust to structural perturbations.

\end{document}